\documentclass[12pt]{article}

 \usepackage{amssymb}
 \usepackage{amsmath}
 \usepackage{amsfonts}
 \usepackage{color}
\newcommand{\be}{\begin{equation}}
\newcommand{\ee}{\end{equation}}
\newcommand{\bea}{\begin{eqnarray}}
\newcommand{\eea}{\end{eqnarray}}
\newcommand{\nn}{\nonumber}
\newcommand{\n}{\nabla}
\newcommand{\rd}{\partial}

\begin{document}

 \begin{titlepage}
  \thispagestyle{empty}

  \vspace{2cm}

  \begin{center}
    \font\titlerm=cmr10 scaled\magstep4
    \font\titlei=cmmi10 scaled\magstep4
    \font\titleis=cmmi7 scaled\magstep4
     \centerline{\titlerm Form field theory on Calabi-Yau threefold }

    \vspace{1.5cm}
    \noindent{{
       Farhang Loran\footnote{e-mail: loran@cc.iut.ac.ir},
        Hesam Soltanpanahi\footnote{e-mail: h\_soltanpanahi@ph.iut.ac.ir}
         }}\\
    \vspace{0.8cm}

   {\it Department of  Physics, Isfahan University of Technology,  \\
  Isfahan 84156-83111,  Iran}

  \end{center}

  \vskip 2em

  \begin{abstract}
   We show how the natural Abelian duality of 2 and 3-form gravity theories on seven-dimensional manifold
   $CY_3\times S^1$ leads to an S-duality between 2 and 3-form theories on simply connected $CY_3$.
  The massless sector of the 2-form field theory on $CY_3$ corresponds to the  A model string field theory.
  We discuss the complex structure independence of the 2 form theory for a general K\"{a}hler manifold
  and derive the holomorphic anomaly equations for simply connected
  $CY_3$.
  \end{abstract}

\end{titlepage}

 \tableofcontents



 \section{Introduction and conclusion}
Since the beginning of 90s, form theories of gravity have been
studied by several authors. Interesting examples of them, related to
the present work are the Chern-Simons theory exposition of
3-dimensional gravity \cite{Witten}, the description of the target
space theory of A model topological string  in terms of variations
of the K\"{a}hler 2-form \cite{sadov}, and the description of the
target space of B model topological string theory in terms of
variations of the holomorphic 3-form \cite{Vafa}-\cite{Loran}.

 In \cite{mirror} Vafa conjectured that there is a mirror symmetry
 between topological A and B string theories. For
 topological string theories targeting a simply connected
 Calabi-Yau threefold, the observables of the A model are in one
 to one correspondence with the elements of $H^{1,1}(CY_3)$ while
 they are independent of the complex structure moduli.
 On the other
 hand, for the topological B model the observables are in one to
 one correspondence with the elements of $H^{2,1}(CY_3)$ but
 independent of the K\"{a}hler structure. The
 existence of such theories is supported by the Candelas
 observation
 in \cite{Candelas} that the moduli space of $CY_3$
 factorizes into a direct product of complex structure moduli and K\"{a}hler moduli at least locally.
 For a simply connected
 Calabi-Yau threefold, $h^{1,1}$ and $h^{2,1}$ i.e. the dimensions of $H^{1,1}$ and $H^{2,1}$ respectively are the only nontrivial
 independent Hodge numbers. A Calabi-Yau threefold mirror to a
 given Calabi-Yau threefold is identified with Hodge numbers,
 \be
 {\tilde h}^{1,1}=h^{2,1},\hspace{1cm}{\tilde h}^{2,1}=h^{1,1}.
 \ee
 The conjectured mirror symmetry of \cite{mirror} is a duality
 between the A model on a $CY_3$ and the B model on the
 corresponding mirror which has been confirmed for example for quintics and the corresponding mirrors. See e.g. \cite{Mirror Symmetry}
 and references therein.

 Roughly speaking the relation between the A model and B model topological
 string theories resembles the relation between type IIA and IIB
 superstring theories. The T-duality between type IIA and IIB superstring theories is explained in terms of an M-theory
 defined on the superstring target space trivially fibred over an extra Euclidean $S^1$.
 Therefore it is
 natural to expect that there exists a
 topological M-theory in 7 dimensions which can be used to
 describe the
 duality of topological string field theories. Also, one expects that
 this 7-dimensional theory is defined on $CY_3\times S^1$, where
 $CY_3$ is trivially fibred over $S^1$.
 The trivially fibred $CY_3\times S^1$ is one of the simplest examples of $G_2$  manifolds
 which are the natural classes of 7-dimensional manifolds  preserving
 some supersymmetry and are purely geometric \cite{Hitchin}.
 This is the setup of \cite{td} where,  a duality of A and B model string field theories on the same target
 space is studied. In principle, the mirror symmetry of the A and B string field
 theories can be given by a topological M-theory for which the
 $CY_3$ is non trivially fibred over $S^1/\mathbb{Z}_2$ which
 fixed points are the locations of a Calabi-Yau threefold and the
 corresponding mirror.

 The topological A and B string field theories can be considered   as form gravity
 theories in 6 or 7 dimensions.
 Recently, in \cite{s1,s2} it is conjectured that there is an S-duality between
 A and B models targeting the same $CY_3$. They
 have  shown that the perturbative amplitudes of the B model
 capture the non-perturbative amplitudes  of the A model and vice
 versa.
 In \cite{td} it is demonstrated that there is a 7-dimensional
 form theory on $CY_3\times S^1$, called topological M-theory, which is related to both
 2 and 3-form theories on the same $CY_3$, and at least at the
 classical level these form theories are related to the topological A
 and B models respectively. Also, they have deduced a
 position/momentum duality, called S-duality, between the
 the 2 and 3-form theories.
 They have explained that the S-duality, indeed, is a duality of A and B$+\bar {\mbox{B}}$ models.
 An evidence for that observation is provided by the BPS counting
 of black hole states \cite{td}. In the form field theories which we are
 going to discuss in the present work, one encounters a similar
 duality of A and B$+\bar {\mbox{B}}$ models but in the case of
 non-simply connected Calabi-Yau threefolds.

 The 7-dimensional manifolds have a specific property which makes
 them useful to be used in a description of the
 A and B models in terms of an M-theory. From
 Abelian duality we know that on every $m$-dimensional manifold $M$
 there is a duality between $r$-forms and $(m-r-2)$-forms \cite{ad}.
 Therefore on a 7-dimensional manifold there is a natural Abelian duality between 2 and 3-form field theories,
 while  in 6 dimensions there does not exist an Abelian duality between 2
 and 3-form fields.
 In this paper we consider the Abelian duality between 2 and 3-form field theories on a trivially fibred $CY_3\times
 S^1$. By applying the usual dimensional reduction method we
 obtain 2 and 3-form field theories on $CY_3$ which are dual to each other in
 the sense of \cite{td}.

 The physical quantities of the form field theories
 we consider in this paper are by construction  elements of $H^{1,1}(CY_3)$,
 since we are considering a simply connected Calabi-Yau threefold ($h^{1,0}=h^{2,0}=0$),
 and $H^3(CY_3)=H^{3,0}\oplus H^{2,1}\oplus H^{1,2}\oplus
 H^{0,3}$. The theory by construction is
 independent of the complex structure. But as can be seen from the
 foregoing Hodge decomposition any representation of the
 observables amounts to defining a complex structure. Therefore,
 the physical observables satisfy a set of identities imposed by  background independence
 conditions which ensure the independence of the complex
 structure \cite{Witten,Verlinde,Loran}. For the 2-form theory, background independence is
 reasonable if the 2-form field theory  is going to be interpreted as the A model string field
 theory. As we will show the massless sector of the 2-form theory we
 obtain by dimensional reduction is the theory of K\"{a}hler
 deformation of the Calabi-Yau threefold. To our knowledge such a theory has been proposed
 for the first time in \cite{sadov} where a theory of K\"{a}hler
 deformation was constructed in analogy to the Kodaira-Spencer theory of complex structure deformation, following the mirror symmetry.
 We show that our proposal for the 2-form theory coincides with
 the AKS theory of \cite{sadov} if all terms in the perturbation
 theory are considered. The 3-form theory studied here is not related to the KS
 theory since it is the Abelian dual  and not the mirror of  the 2-form theory and is defined on
 the same Calabi-Yau threefold and not on its mirror threefold.
 We further explain this point in terms of a classic/quantum
 correspondence between the two form field theories.

 In this paper we study duality of 2 and 3-form theories which at
 the classical level correspond to the topological A and B string
 field theories on the same $CY_3$ respectively. To this aim we use
 Abelian duality in 7 dimensions. We explicitly show that the
 Abelian duality of the 2 and 3-form gravity theories on a trivially fibred $CY_3\times S^1$ gives, after Kaluza-Klein reduction,
 an S-duality between 2 and 3-form
 gravity theories on $CY_3$.
 From a 7 dimensional point of view the Abelian duality gives the position/momentum duality of
 \cite{td}. Furthermore both in 6 and 7 dimensions the Abelian duality implies a classic/quantum correspondence  between the two theories.
 In fact, the classical regime of the 2-form theory corresponds to the
 3-form theory in the $\hbar\rightarrow\infty$ limit.
 In the classical limit, the 2-form theory reduces to its zero mode
 sector which is the theory of the K\"{a}hler structure deformation.
 The other extreme is the classical regime of the 3-form field
 theory which corresponds to the $\hbar\rightarrow\infty$ limit of
 the 2-form theory. The 3-form theory in this limit reduces to its
 zero mode sector in which the physical quantities are the elements of
 $H^3(CY_3)$.
 It is tempting to anticipate a connection between this
 theory and the topological B string field theory. But, we notify again
 that, the 3-form theory should not be expected to be the
 Kodaira-Spencer theory since one is considering a trivially
 fibred $CY_3$ over $S^1$.

 In  a proper gauge we  quantize these
 form gravity theories and describe the Abelian duality as a map between the corresponding annihilation and creation operators.
 An interesting feature of these theories is the existence of a one parameter family of
 vacua,
 \begin{equation}
 \left|\textrm{vac.}\right>_\theta=e^{i\theta a_0}\left|0\right>,
 \end{equation}
 where $a_0$ corresponds to the massless modes.
 The quantum theory of 2-form field theory should be further
 equipped by suitable holomorphic anomaly equations ensuring the
 background independence
 since this is  a gravity theory for the A
 model. On the dual side, quantum background independence  for
 3-form fields is previously worked out in \cite{Verlinde,Gerasimov,Loran}.

The plan of this paper is as follows: in section 2, after a brief
review of Abelian duality on general $m$-dimensional manifolds we
study Abelian duality in the case of the 7-dimensional manifold
$CY_3\times S^1$ between 2 and 3-form theories. In a proper gauge
the equations of motion of the original field theory reduce to the
equations for components of fields along $CY_3$. This makes the
dimensional reduction of 7-dimensional form field theories to 6
dimensions straightforward. We show that classically the Abelian
duality is a position/momentum duality. In the path integral
formulation of the Abelian duality  we show that it describes a
classic/quantum correspondence between the two form field
theories. In section 3, we derive 2 and 3-form gravity theories on
$CY_3$ by applying the KK dimensional reduction method to the
original 7-dimensional form field theories and reinterpret the
classic/quantum correspondence in 6 dimensions. Canonical
quantization of the 7-dimensional field theories is studied in
section 4. Section 5 is devoted to the connection between the
2-form theory and the theory of K\"{ahler} gravity. In section 6
we first construct a method to study the complex structure
deformation of a complex manifold. Then we study the quantum
background independence of the massless sector of the 2-form
theory. In appendix $A$ we derive equations used in section 2. The
mathematical background of section 3 is provided in appendix $B$.

 \section{Abelian duality in 7 dimensions}\label{sec 2}
In this section we  derive Abelian duality between 2 and 3-forms
in 7 dimensions. For doing this in  subsection \ref{sec 2.1} we
study Abelian duality of $r$ and $(m-r-2)$-forms on an
$m$-dimensional manifold $M$. Then in subsection \ref{2.2} we use
Abelian duality in the 7-dimensional manifold to explain relation
between 2 and 3-form theories in 7 dimensions. In subsection
\ref{2.3}, for the special 7-dimensional manifold $CY_3\times S^1$
by using gauge fixing condition and after decomposing the
components of 2 and 3-forms along $CY_3$ and $S^1$ we obtain
equations of motion for components of forms along $CY_3$ and the
classical duality between those components. The result is a
duality between the 2 and 3-form field theories in 6 dimensions,
which is  the inheritance of the 7-dimensional Abelian duality.
\subsection{Abelian duality}\label{sec 2.1}
In this subsection we review Abelian duality in $m$-dimensional
Riemannian manifolds $M$ and define  form field theories on such
manifolds. All of the operators which one encounters in our form
theories are shown in the following diagram \cite{ad}
 \bea
\begin{picture}(300, 100)\thinlines
\put(105, 75){\parbox{45pt}{$\Omega^r(M)$}} \put(152,
79){\vector(1,0){50}} \put(212,
75){\parbox{50pt}{$\Omega^{r+1}(M)$}} \put(118,
70){\vector(0,-1){30}} \put(230, 70){\vector(0,-1){30}} \put(100,
28){\parbox{50pt}{$\Omega^{m-r}(M)$}} \put(152,
32){\vector(1,0){50}} \put(208,
28){\parbox{63pt}{$\Omega^{m-r-1}(M)$}} \put(170,
38){\parbox{50pt}{$\scriptstyle d^{\dagger}$}} \put(170,
82){\parbox{50pt}{$\scriptstyle d$}} \put(122,
52){\parbox{10pt}{$\scriptstyle \ast$}} \put(234,
52){\parbox{10pt}{$\scriptstyle \ast$}}
\end{picture}\nn
\eea
 where $\Omega^r(M)$ is the set of  $r$-forms on M. If $M$ is a Riemannian manifold, a general
 $r$-form $\eta_r$ satisfies the following equations:
 \begin{eqnarray}
&&d^\dagger\eta_r=(-)^{m(r+1)+1}\ast d \ast\eta_r,\label{dag1}\\
&&\ast\ast\eta_r=(-)^{r(m-r)}\eta_r,\label{ast}
\end{eqnarray}
where $d$ is the exterior derivative operator and $*$ is the Hodge
star operator. There are the same relations with an extra $(-1)$
for Lorentzian
 manifolds.
One can define an inner product of two $r$-forms $\eta_r$ and
$\zeta_r$ on a  manifold $M$,
\begin{equation}
(\eta_r , \zeta_r)=\int_{_M}\eta_r\wedge\ast\zeta_r,
\end{equation}
which is positive definite  for Riemannian manifold,
${\left(\varpi_r , \varpi_r\right)\geq 0}$, and the
 equality is satisfied if and only if $\varpi_r=0$. Furthermore,
 $\left(\eta_r , \zeta_r\right)=\left(\zeta_r , \eta_r
\right)$ \cite{Nakahara}.

For a closed $r$-form $\eta_r$ which obeys the condition
$d\eta_r=0$ one can locally assume that $\eta_r=d\zeta_{r-1}$.
This follows from the fact that the operator $d$ is nilpotent
namely $d^2=0$. Also for the nilpotent operator $d^\dag$    if
$d^\dag\eta_r=0$ then locally $\eta_r=d^\dag\zeta_{r+1}$.
Therefore if
\begin{equation}
d^\dag (d\gamma_r)=(-)^{m(r+2)+1}\ast d\ast
d\gamma_r=0,\label{dag}
\end{equation}
in which  we have used Eq. (\ref{dag1}), then by using Eq.
(\ref{ast}), one obtains
\begin{equation}
d(\ast d\gamma_r)=0.
\end{equation}
Thus there exists an $(m-r-2)$-form $\omega_{m-r-2}$ for which
\begin{equation}
\ast d\gamma_r=d\omega_{m-r-2}.\label{ab}
\end{equation}
This relation is known as \textit{Abelian duality}. Considering
both the $r$-from $\gamma_r$ and the ${(m-r-2)\mbox{-form}}$
$\omega_{m-r-2}$ as classical fields of two different form field
theories, then Eq. (\ref{ab}) is a duality of the corresponding
equations of motions.

 By using a positive definite inner product we can
construct an action for any ${r\mbox{-form}}$ $\eta_r$ such that
it gives an equation of motion for $\eta_r$ similar to Eq.
(\ref{dag}). This action is defined in the following way:
\begin{equation}
S(\eta)\sim(d\eta_r , d\eta_r)=\int_{_M}d\eta_r\wedge\ast d\eta_r,
\end{equation}
for which, using the identity,
 \be
 (d\alpha_r,\beta_r)=(\alpha_r,d^\dag\beta_r),
 \ee
 the equation of motion is given by,
\begin{equation}
d^\dag d\eta_r=0.\label{e}
\end{equation}
Also, it is obvious that this action is invariant under the
following gauge transformation,
\begin{equation}
\eta_r\rightarrow\eta_r+d\theta_{r-1},
\end{equation}
where $\theta_{r-1}\in\Omega^{r-1}(M)$. The gauge freedom can be
fixed  by using the following gauge condition,
\begin{equation}
d^\dag\eta_r=0.\label{g}
\end{equation}
By using both equation of motion (\ref{e}) and gauge fixing
condition, (\ref{g}) one obtains that\footnote{By definition on
$m$-dimensional manifold $\Delta=d^\dag d+d^\dag d$.}
\begin{equation}
\Delta\eta_r=0.\label{dalam}
\end{equation}
 Eq. (\ref{dalam}) implies that the gauge (\ref{g}) does not fix the gauge freedom completely since one can still add an arbitrary harmonic form to $\eta$.
 \subsection{7-dimensional Abelian duality}\label{2.2}
 Now, we are ready to study Abelian duality of 2 and 3-form field theories on $\Sigma=CY_3\times S^1$.
 Considering $C\in \Omega^3(\Sigma)$ and $K\in \Omega^2(\Sigma)$, we assume that both of these fields obey the
 following equations of motion\footnote{Subscript of operators $d$ and $*$ refers to
 the dimensions of manifolds on which the operators are defined.}
 \begin{equation}
 d^\dag_7d_7 C=0,\hspace{2cm}d^\dag_7d_7 K=0,\label{eom}
 \end{equation}
with the Abelian duality relation
\begin{equation}
 \ast_7d_7C=d_7K.\label{cd}
\end{equation}
To study the quantum version of this classical duality one needs
to compare the partition functions of both theories and
consequently one should define an action functional for each of
them. As we saw in the last subsection, on a compact orientable
Riemannian  manifold one can define a positive definite action for
the 3-form $C$,
\begin{equation}
 S(C)=\frac{\Lambda}{4\pi}\left(d_7C , d_7C\right)=\frac{\Lambda}{4\pi}\int_\Sigma d_7C\wedge\ast_7d_7C,\label{C action}
\end{equation}
which gives Eq. (\ref{eom}) for the 3-form $C$. The partition
function of theory is,
\begin{equation}
 Z=\int\mathcal{D}\phi e^{-S(\phi)}.
\end{equation}
To obtain the Abelian duality, following \cite{duality}, we
introduce
 new variables such that if we
integrate them out from the partition function we get back to the
original theory, and if we integrate out the old variables
instead of the new ones it produces the dual  theory. To this aim
we interpret $C$ as a section of a trivial $S^1$-bundle
$\mathcal{L}_A$ with a connection $A$, and write an action which
includes the covariant derivative $D_{7A}C=d_7C+A$ in the
following way,
\begin{equation}
 S(C,A)=\frac{\Lambda}{4\pi}\int_\Sigma D_{7A}C\wedge\ast_7D_{7A}C.\label{CA action}
\end{equation}
There is a simple mechanism for recovering the old action. We
introduce a third field $K$ which plays the role of the Lagrange
multiplier. We take $K$ to be a section from $\Sigma$ to $S^1$,
and write an action,
\begin{equation}
 S(C,A,K)=\frac{\Lambda}{4\pi}\int_7D_{7A}C\wedge\ast_7D_{7A}C-\frac{i}{2\pi}\int_7K\wedge d_7A.\label{CAK action}
\end{equation}
 This action is invariant under the gauge transformation
 \bea
 &&A\rightarrow A+dB,\nn\\
 &&C\rightarrow C-B,\label{gfc}
 \eea
 where B is a 3-form.
 To see that $S(C,A,K)$ is equivalent to $S(C)$, one can note
 that the equation of motion for $K$ is $d_7A=0$. By imposing this
 condition and using the $A=0$ gauge the original theory gets recovered.

 The above naive argument can be made exact as follows. Consider
 the
 partition function of the whole theory,
 \begin{equation}
 Z=\frac{1}{vol(G)}\int \mathcal{D}C\mathcal{D}A\mathcal{D}K
 \exp{\left[-\frac{\Lambda}{4\pi}\int_7D_{7A}C\wedge\ast_7D_{7A}C+\frac{i}{2\pi}\int_7K\wedge d_7A\right]},\label{CAK pi}
 \end{equation}
 in which $G$ is the gauge group for $\mathcal{L}_A$. By using the
 delta function,
 \begin{equation}
 \int\frac{d x}{2\pi} e^{ixy}=\delta(y),
 \end{equation}
 the $K$ integral  sets $dA$ equal to zero. But we are
studying circle-valued forms and we should treat this statement
carefully. In general $K=K_{har}+K_{S^1}$, where
$K_{har}:\Sigma\rightarrow S^1$ is a circle valued 2-form such that
$dK_{har}$ is a harmonic 3-form representative in the cohomology
class of $dK$, and $K_{S^1}$ is a single valued real function on
$S^1$. Introducing a basis for harmonic 3-forms $\lambda_j$, where
$j=1,\dots ,h^{3}(\Sigma)$,\footnote{
$h^{3}(\Sigma)=2+2h^{2,1}(CY_3)+h^{1,1}(CY_3)$\cite{Nakahara}.} one
can write $dK_{har}= \sum_{j=1}^{h^3(\Sigma)} 2\pi m_j\lambda_j$, in
which $m_j\in \mathbb{Z}$. Now one obtains,
\begin{eqnarray}
&&\int\mathcal{D}K \exp\left[\frac{i}{2\pi}\int_\Sigma K\wedge d_7A\right]\nn\\
&&=\int\mathcal{D}K_{S^1} \exp\left[\frac{i}{2\pi}\int_\Sigma
K_{S^1}\wedge d_7A\right]\times
\sum_{dK_{har}\in H^3(\Sigma,2\pi \mathbb{Z})}e^{-\frac{i}{2\pi}dK_{har}\wedge A}\nn\\
&&=\delta(d_7A)\prod_j\left(\sum_{m_j\in
\mathbb{Z}}e^{-im_j\int_\Sigma\lambda_j\wedge A}\right).
\end{eqnarray}
The first term shows that $A$ is a flat connection, and among flat
connections the gauge equivalence classes $[A]$ are labeled by
holonomies, or equivalently by the quantities
$\int_{\Sigma}\lambda_j\wedge A$. The remaining part of the
integral is gauge invariant therefore  one can fix the gauge and
integrate over the space of gauge equivalent classes and obtain
(omitting the factor$(vol(G))^{-1}$),
\begin{eqnarray}_{}
 Z&=&\int \mathcal{D}C \mathcal{D}[A]e^{-\frac{\Lambda}{4\pi}\int_\Sigma D_{7A}C\wedge\ast_7D_{7A}C}
 \delta(d_7A)\prod_j\left(\sum_{m_j\in \mathbb{Z}}e^{-im_j\int_\Sigma\lambda_j\wedge A}\right)\nn\\
 &=&\int \mathcal{D}C e^{-\frac{\Lambda}{4\pi}\int_\Sigma D_{7A}C\wedge\ast_7D_{7A}C}
 \delta(d_7A)\delta\left(\int_\Sigma\lambda_j\wedge A=0\hspace{4mm}\textrm{mod}\hspace{2mm}2\pi
 \mathbb{Z}\right)\nn\\
 &=&\int\mathcal{D}C e^{-\frac{\Lambda}{4\pi}\int_\Sigma d_{7}C\wedge\ast_7d_{7}C}.\label{C[A] pi}
\end{eqnarray}
In the last equality we have used the fact that the delta
functions sets the holonomies equal to zero as well as curvature.
Thus $A$ is zero modulo gauge transformations, and the new theory
is indeed equivalent to the original theory.

Now we want to integrate out $C$ and $A$ to obtain a new theory
in terms of the 2-form field $K$. First, we integrate out $C$ by
using the Faddeev-Popov method and then integrate $A$ out by
Gausian method. For the first step, we fix the  gauge freedom
(\ref{gfc}) using the gauge $C=0$. In this gauge  the path
integral (\ref{CAK pi}) reduces to,
\begin{equation}
 \int \mathcal{D}A \mathcal{D}K \mathcal{D}\overline{c} \mathcal{D}c
 \exp\left[-\frac{\Lambda}{4\pi}\int_\Sigma A\wedge\ast_7A+\frac{i}{2\pi}\int_\Sigma K\wedge d_7A
  -\frac{\Lambda}{4\pi}\int_\Sigma \overline{c}1c\right].\label{AK pi}
\end{equation}
The last term arises from gauge fixing by Faddeev-Popov method,
and $c$ and $\overline{c}$ are any allowed grassmanian forms on
$\Sigma$.

For integrating out $A$, we complete the square. To do this, we
introduce new variable $A'=A+\frac{i}{\Lambda}\ast_7d_7K$. Then
we have,
\begin{equation}
\left(\int \mathcal{D}K e^{-\frac{1}{4\pi\Lambda}\int_\Sigma
dK\wedge\ast_7d_7K}\right) \left( \int \mathcal{D}A'
e^{-\frac{\Lambda}{4\pi}\int_\Sigma A'\wedge\ast_7A'}\right)
\left(\int \mathcal{D}\overline{c} \mathcal{D}c
e^{-\frac{\Lambda}{4\pi}\int_\Sigma \overline{c}1c}\right).
\end{equation}
The first term is the partition function of the dual theory which
is the 2-form theory in 7 dimensions. The  last two terms give a
topological factor which depends on the number of independent
harmonic forms in 7-dimensional manifold $CY_3\times S^1$ (see
appendix A for this subtlety). Consequently we have the following
equality between the partition functions of  two form theories,
\begin{equation}
\int\mathcal{D}C e^{-\frac{\Lambda}{4\pi}\int_\Sigma
dC\wedge\ast_7d_7C}=\mathcal{C}(\Sigma)\int \mathcal{D}K
e^{-\frac{1}{4\pi\Lambda}\int_\Sigma
dK\wedge\ast_7d_7K},\label{qd}
\end{equation}
It is notable that the coupling $\Lambda$ in the first theory is
mapped to $1/\Lambda$ in the second theory. Therefore there is a
classic/quantum correspondence between these form theories.
Equation (\ref{qd}) is given in the in the natural units
($c=\hbar=1$). Recovering $\hbar$ in Eq. (\ref{qd}) one obtains
$Z(\phi)=\int\mathcal{D}\phi e^{-\frac{1}{\hbar}S(\phi)}$. Thus,
in 3-form (2-from) theory $1/\Lambda$ ($\Lambda$) acts as $\hbar$
and we can study classical and quantum regimes of each theory in
the limit of $\Lambda$ tending to $0$ or $\infty$.
 \subsection{Equation of motion and gauge fixing}\label{2.3}
 The equation of motion for 3-form $C$ given by the action (\ref{C action})
 is,
\begin{equation}
d^\dag_7 d_7 C=0.\label{eom C}
\end{equation}
After the decomposition $C=\gamma+dt\wedge\alpha$ where $\gamma$
is a 3-form  along $CY_3$ and $dt\wedge\alpha$ is a 1-form along
$S^1$ and a 2-form along $CY_3$, and also the decomposition of the
operator $d_7$ into its components along $CY_3$ and $S^1$,
$d_7=d_{CY}+dt\wedge d/dt$, the equation of motion (\ref{eom C})
reduces to, \footnote{We can obtain these equations of motion by
putting decomposition of $C$ in action(\ref{C action}) and using
Euler-Lagrange equation. }
\begin{eqnarray}
d^\dag_{_{CY}}d_{_{CY}}\gamma-\ddot{\gamma}-d_{_{CY}}\dot{\alpha}&=&0,\label{eom gamma}\\
d^\dag_{_{CY}}d_{_{CY}}\alpha+d^\dag_{_{CY}}\dot{\gamma}&=&0,\label{eom
alpha}
\end{eqnarray}
where dot $``."$ stands for a derivation with respect to $t$ which
parameterizes the $S^1$. To obtain this results one needs the
relations between operators $d$ and $\ast$ in the 7-dimensional
manifold $\Sigma$ and the 6-dimensional manifold $CY_3$ which can
be found in appendix B.

The action (\ref{C action})  is invariant under the gauge
transformation $C\rightarrow C+d_7\rho$, where $\rho$ is a 2-form
along $\Sigma$. We can fix this gauge freedom by $d^\dag_7C=0$
gauge. In terms of $\gamma$ and $\alpha$, the gauge
transformation is,
 \be
 \gamma\to\gamma+d\rho_1,\hspace{1cm}\alpha\to\alpha+\dot{\rho}_1+d\rho_2,
 \label{rho}
 \ee
 in which $\rho=\rho_1+\rho_2\wedge dt$ and
 the gauge fixing condition is
equivalent to,
\begin{eqnarray}
d^\dag_{_{CY}}\gamma+\dot{\alpha}&=&0,\label{gf gamma} \\
d^\dag_{_{CY}}\alpha&=&0.\label{gf alpha}
\end{eqnarray}
By operating $d_{CY}$ and $d/dt$  on the first equation and
inserting the result into the  equations of motion (\ref{eom
gamma},\ref{eom alpha}), one obtains,
\begin{eqnarray}
\Delta_{_{CY}}\gamma-\ddot{\gamma}&=&0,\label{eomg gamma}\\
\Delta_{_{CY}}\alpha-\ddot{\alpha}&=&0.\label{eomg alpha}
\end{eqnarray}
Similar calculations can be done in the case of the  2-form theory
in 7 dimensions. From Eq. (\ref{qd}) one verifies  that the action
for $K$ is,
\begin{equation}
S(K)=\frac{1}{4\pi\Lambda}\int_{\Sigma}d_7K\wedge\ast_7d_7K.\label{K
action}
\end{equation}
After the decomposition $K=\omega+dt\wedge\beta$, where $\omega$
and $\beta$ are 2-form and 1-form along $CY_3$ respectively, the
equation of motion for $K$, i.e. $d^\dag_7 d_7 K=0$, reduces to,
\begin{eqnarray}
d^\dag_{_{CY}}d_{_{CY}}\omega-\ddot{\omega}+d_{_{CY}}\dot{\beta}&=&0,\label{eom omega}\\
d^\dag_{_{CY}}d_{_{CY}}\beta-d^\dag_{_{CY}}\dot{\omega}&=&0.\label{eom
beta}
\end{eqnarray}
The gauge symmetry of  the action for 2-form $K$ is,
\begin{equation}
K\rightarrow K+d\theta,
\end{equation}
where $\theta$ is an 1-form along $\Sigma$. The gauge fixing
condition $d^\dag_7K=0$ gives,
\begin{eqnarray}
d^\dag_{_{CY}}\omega-\dot{\beta}&=&0,\\
d^\dag_{_{CY}}\beta&=&0,
\end{eqnarray}
which can be used to simplify the equations of motion (\ref{eom
omega},\ref{eom beta}) as follows,
\begin{eqnarray}
\Delta_{_{CY}}\omega-\ddot{\omega}&=&0,\\
\Delta_{_{CY}}\beta-\ddot{\beta}&=&0.
\end{eqnarray}

The classical duality (\ref{cd}) between the 3-form $C$ and 2 form
$K$ results in the following duality relations,
\begin{eqnarray}
&&\ast_{_{CY}}d_{_{CY}}\gamma=\dot{\omega}+d_{_{CY}}\beta,\label{cd gamma}\\
&&\ast_{_{CY}}(\dot{\gamma}+d_{_{CY}}\alpha)=d_{_{CY}}\omega.\label{cd
alpha}
\end{eqnarray}
In the next sections we will argue that these relations give us
an interpretation of the Abelian duality as the position/momentum
duality \cite{td}.

The 3-form action (\ref{C action}) in terms of the $\gamma$ and
$\alpha$ components is,\footnote{Henceforth, all operators and the
corresponding inner products are defined on $CY_3$ and we omit
the subscript $CY$.}
\begin{equation}
 S(\gamma,\alpha)=\frac{\Lambda}{4\pi}\int_{S^1}dt\left\{(\dot{\gamma},\dot{\gamma})+(d\gamma,d\gamma)+
 2(\dot{\gamma},d\alpha)+(d\alpha,d\alpha)\right\}.
 \label{gamma,alpha action}
\end{equation}
Since we are interested in form field theories on $CY_3$, we use
the gauge freedom (\ref{rho}) to set,
 \be
 \alpha=0.
 \ee
In this gauge the action (\ref{gamma,alpha action}) simplifies to,
\be
 S(\gamma)=\frac{\Lambda}{4\pi}\int_{S^1}dt\left\{(\dot{\gamma},\dot{\gamma})+(d\gamma,d\gamma)\right\}.
 \label{decouple}
\ee One should note that there is still a gauge freedom,
 \be
 \gamma\to\gamma +d{\tilde\rho}_1,
 \ee
 where ${\tilde\rho}_1$ is an arbitrary 2-form independent of $t$,
 \be
 {\tilde\rho}_1\in\Omega^2(CY_3).
 \label{tilde rho}
 \ee

 By the same argument one can get
rid of the 1-form $\beta$ in the 2-from theory to obtain
$S(\omega)=\frac{1}{4\pi\Lambda}\int_{S^1}dt\left\{(\dot{\omega},\dot{\omega})+(d\omega,d\omega)\right\}$.
In these gauges the classical dualities (\ref{cd gamma}) and
(\ref{cd alpha}) reduce to,
\begin{eqnarray}
&&*d\gamma=\dot{\omega},\label{d gamma}\\
&&*\dot{\gamma}= d\omega.\label{d alpha}
\end{eqnarray}
 \section{Reduction to Calabi-Yau threefold}\label{3}
 In this section we study the reduction of 2 and 3-form
theories in 7 dimensions to form gravity theories on 6-dimensional
manifold $CY_3$ by using Kaluza-Klein method. We assume that $R$
is the radius of $S^1$ over which the $CY_3$ is fibred trivially.
In this way we can expand all forms,
\begin{eqnarray}
&&\gamma=\sum_{n\in \mathbb{Z}}\gamma_ne^{i2\pi nt/R},\\
&&\omega=\sum_{n\in \mathbb{Z}}\omega_ne^{i2\pi nt/R},\label{kk
mode}
\end{eqnarray}
where $\gamma_n$ and $\omega_n$ are defined on $CY_3$ and their
dependence on the extra dimension $t$ along $S^1$ is given in the
phase $e^{i2\pi nt/R}$.
 Thus both actions on the $CY_3$ reduce to,
\begin{eqnarray}
&&S(\gamma_n)=\frac{\Lambda}{4\pi}\sum_{n\in
\mathbb{Z}}\left\{\frac{4\pi^2n^2}{R^2}(\gamma_n,\gamma_{-n})+
(d\gamma_n, d\gamma_{-n})\right\},\label{gamma action}\\
&&S(\omega_n)=\frac{1}{4\pi\Lambda}\sum_{n\in
\mathbb{Z}}\left\{\frac{4\pi^2n^2}{R^2}(\omega_n,\omega_{-n})+
(d\omega_n, d\omega_{-n})\right\}.\label{omega action}
\end{eqnarray}
 To obtain the above results the following identity is used,
\begin{equation}
\int_{S^1}dt e^{i2\pi(m+n)t/R}=\delta_{m+n,0}
\end{equation}
 In addition, all of the equations of motion on $CY_3$ in
terms of KK modes simplify to,
\begin{eqnarray}
&&d^\dag d\gamma_n=-\frac{4\pi^2n^2}{R^2}\gamma_n,\label{eom gamma n}\\
&&d^\dag d\omega_n=-\frac{4\pi^2n^2}{R^2}\omega_n.\label{eom omega
n}
\end{eqnarray}
 We see that coefficient the $\frac{4\pi^2n^2}{R^2}$ in both theories gives the KK mass
squared of the $n$-th mode.

 There is a subtlety  in the partition functions of these theories as can be seen from Eqs.(\ref{gamma action},\ref{omega action}).
 In the classical limit of the 3-form (2-form) theory  $\Lambda\rightarrow \infty$ ($1/\Lambda\rightarrow \infty$),
 which corresponds to  $\hbar\rightarrow 0$,
 the partition function  gets localized around the classical trajectory.
 By using equations of motion (\ref{eom gamma n},\ref{eom omega n})
 one can obtains the following inner product for the KK modes on the Calabi-Yau
 threefold
 \bea
 \left(d\gamma_n , d\gamma_n\right)&=&-\frac{4\pi^2n^2}{R^2} \left(\gamma_n , \gamma_n\right),\nn\\
 \left(d\omega_n , d\omega_n\right)&=&-\frac{4\pi^2n^2}{R^2} \left(\omega_n , \omega_n\right).
 \eea
 Recalling that for Riemannian manifolds  the inner product is positive definite,
 one verifies that the solution of the equations of motion is $\gamma_n=0=\omega_n$ for $n\neq 0$.
 Consequently in the classical limit each theory reduces to its
 zero mode sector. Furthermore, one easily verifies the
 classic/quantum correspondence between the two form field
 theories is a correspondence between the massless sector of one
 theory and the full spectrum of the another one. In section \ref{5} we will show
 that the massless sector of the 2-form theory gives the theory
 of K\"{a}hler gravity on the $CY_3$. Thus the classic/quantum
 correspondence implies why the massless sector of the 3-form field
 theory is not the Kodaira-Spencer theory which is believed to correspond to
 the topological B string field theory.


 \section{Canonical quantization and position/momentum duality}\label{sec 4}
 Now we are able to do canonical quantization. Recall that the actions
 of 3 and 2-form theories in 7 dimensions respectively are,
 \begin{eqnarray}
 && S(\gamma)=\frac{\Lambda}{4\pi}\int_{S^1}dt\left\{(\dot{\gamma},\dot{\gamma})+(d\gamma, d\gamma)\right\},\label{1}\\
 &&S(\omega)=\frac{1}{4\pi\Lambda}\int_{S^1}dt\left\{(\dot{\omega},\dot{\omega})+(d\omega,d\omega)\right\}.\label{2}
 \end{eqnarray}
 The momenta conjugate to  $\gamma$ and $\omega$ are respectively,\footnote{Here we have used the following
 definition,
 \begin{equation}
 \frac{\rd(\psi , \chi)}{\rd \psi}=\frac{\rd(\chi , \psi)}{\rd
 \psi}=\chi,
 \end{equation}
  which means that the derivation is defined with respect to the {\em product} $\wedge *$.}
\begin{eqnarray}
&&P_{\gamma}=\frac{\Lambda}{2\pi}\dot{\gamma},\label{m gamma}\\
&&P_{\omega}=\frac{1}{2\pi\Lambda}\dot{\omega}.\label{m omega}
\end{eqnarray}
The  Poisson-bracket is given as usual e.g.
\begin{equation}
\{\gamma , P_\gamma\}=1,
\end{equation}
which results in,
\begin{eqnarray}
&&\{\gamma_n , \gamma_m\}=\frac{i}{n}\frac{R}{\Lambda}\delta_{n,-m},\hspace{1cm} n,m\neq 0,\\
&&\{\gamma_0 , \gamma_n\}=0,\hspace{3cm} n\in\mathbb{Z}.
\end{eqnarray}
One can define the creation and annihilation operators as usual
\begin{eqnarray}
&&a_n=\sqrt{n}\gamma_n,\hspace{15.5mm}n>0,\\
&&a^\dag_{-n}=\sqrt{-n}\gamma_{n},\hspace{1cm}n<0,\\
&&a_0=a^\dag_0=\gamma_0,
\end{eqnarray}
satisfying the algebra
 \begin{eqnarray}
 &&[a_n,a_m^\dag]=\frac{R}{\Lambda}\delta_{n,m},\hspace{1cm}n,m\neq 0,\\
 &&[a_0,a_n]=[a_0,a^\dag_n]=0,\hspace{8mm}n\in\mathbb{Z}.
 \end{eqnarray}
The vacuum state for the 3-form theory $\left|0\right>$ is
defined to be the state annihilated by $a_n$. There is a subtlety
at this point. Since  $a_0$ commutes with the massive operators
$a_n^\dag$, one verifies that this operator
 takes one vacuum state into a nearby equivalent one
\cite{Candelas}. Therefore one can define a one parameter family
of vacua such that,
\begin{equation}
\left|\textrm{vac.}\right>_\theta=e^{i\theta a_0}\left|0\right>.
\end{equation}

We can quantize the 2-form theory (\ref{2}) in the same way and
derive the associated vacua family.  The difference between these
two form field  theories is due to the exponent of $\Lambda$ in
the commutation algebra . In the 2-from theory the commutation
relation is given by,
 \begin{eqnarray}
 &&[b_n,b_m^\dag]=R \Lambda\delta_{n,m},\hspace{1cm}n,m\neq 0,\\
 &&[b_0,b_n]=[b_0,b^\dag_n]=0,\hspace{11mm}n\in\mathbb{Z},
 \end{eqnarray}
where
\begin{eqnarray}
&&b_n=\sqrt{n}\omega_n,\hspace{15.5mm}n>0,\\
&&b^\dag_{-n}=\sqrt{-n}\omega_{n},\hspace{1cm}n<0,\\
&&b_o=b^\dag_0=\omega_0.
\end{eqnarray}
Consequently the classical limit of one theory
$\hbar\rightarrow0$ (e.g. for 3-form theory it means
$\Lambda\rightarrow\infty$) corresponds to the quantum extreme of
the another theory.
 This is the classic/quantum correspondence
of 2 and 3-from theories in $CY_3$ induced by the Abelian duality.


\subsection{S-duality on $CY_3$ }
Now we can understand the duality between the two form gravity
theories.The duality is two-edged indeed. Firstly, there is a
classic/quantum correspondence between two theories. Secondly,
from the classical duality relations (\ref{d gamma},\ref{d
alpha}) and definition of momentum (\ref{m gamma},\ref{m omega})
we obtain the interesting relations,
\begin{eqnarray}
&&\ast d\gamma=2\pi\Lambda P_\omega,\\
&&\ast P_\gamma=\frac{\Lambda}{2\pi}d\omega,
\end{eqnarray}
which can be interpreted as position/momentum or electric/magnetic
 duality in 7-dimensional form theories\cite{td}.
In the 6 dimensional manifold, this relations in terms of KK
modes reduce to,
\begin{eqnarray}
&&\ast d\gamma_n=i\frac{2\pi n}{R}\omega_n,\label{dn gamma}\\
&&\ast d\omega_n=i\frac{R}{2\pi n}\gamma_n.\label{dn alpha}
\end{eqnarray}
There is a subtlety  in these equations. We see that because of
the coefficient $n$, the zero modes of the theories do not play
any role in the above  duality relations. Zero modes play a
significant role in the classical/quantum correspondence. Here we
recall our discussion on the classic/quantum correspondence at
the end of section \ref{3}. In the classical limit, the 3-form
field theory for example, is given by its massless sector which
corresponds to the 2-form theory in its quantum extreme with both
massless and massive excitations.


 \section{ Theory of K\"{a}hler gravity}\label{5}
 Let us see what is the relation between the action,
 \be
 S(\omega)\sim(d\omega,d\omega),
 \ee
  and the theory of K\"{a}hler gravity. It
 is known that the observables of the A-model are elements of
 $H^{(p,q)}$. For $p=q=1$ one verifies a correspondence between
 the A-model and the theory of K\"{a}hler transformations.

 In physical superstring theories we have $N=(2,2)$ supersymmetry.
 After the twist to the topological string theories there remain only
 two supercharges.  In the B string field theory the corresponding
 nilpotent operators are $\rd$ and
 $\bar\rd$. A deformation in the complex structure amounts to
 deformation of $\bar \rd$ given by,
 \be
 \bar\rd\rightarrow\bar\rd+A.\rd,
 \ee
 where $A\in H^{0,1}(TM)$. Requiring that the deformed operator is
 nilpotent one obtains the  Kodaira-Spencer equation\cite{Vafa},
 \be
 \bar\rd A'+\frac{1}{2}\rd(A\wedge A)'=0 ,
 \ee
 where $A'=A.\Omega$ in which $\Omega$ is the Calabi-Yau three form.

 In the A model the nilpotent operators are $d$ and ${d^c}^\dag$.\footnote{By definition we have:
\begin{equation}
d=\rd+\bar\rd,\hspace{10mm}d^c=\bar\rd-\rd.\nn
\end{equation}
One can define $d$ without determining complex structure but
define $d^\dag$ one should choose a complex structure.} Operator
$d$ is nilpotent  without any dependence on the K\"{a}hler and/or
complex structure. Given a K\"{a}hler structure $\omega_0$ the
${d^c}^\dag$ operator is defined as follows,
 \be
  {d^{c}}^\dag=[d,\Lambda_0],
 \label{tt}
 \ee
 where $\Lambda_0$ is the pull back of multiplication by
  $\omega_0$ \cite{Griffiths}. As we see, the definition of the
  operators $d$ and ${d^c}^\dag$ is independent of the complex
  structure, but for ${d^c}^\dag$ it depends on the K\"{a}hler
  structure as is expected for the A string field theory.
  By a K\"{a}hler transformation one means  a
 deformation of the K\"{a}hler form $\omega_0$,
 \be
 \omega_0\to\omega=\omega_0+K,\label{to}
 \ee
 where $K$ is a 2-form, leaving the operator
 ${D^c}^\dag={d^c}^\dag+[d,\delta\Lambda]$  be nilpotent. Again
 $\Lambda=\Lambda_0+\delta\Lambda$ is the pull back of the
 multiplication by $\omega$.
 Given a complex structure, $\Lambda$ in components is
 given by the following identities,
 \bea
 \Lambda^{i\bar j}\omega_{i\bar k}&=&\delta^{\bar j}_{\bar k},\\
 \Lambda^{i\bar j}\omega_{k\bar j}&=&\delta^i_k.
 \eea
 For a general K\"{a}hler form $\omega$ it is straightforward to verify that ${d^c}^\dag=[d,\Lambda]$ is nilpotent if
 and only if,
 \be
 \Lambda^{l\bar j}_{,k}\Lambda^{k\bar m}=\Lambda^{l\bar
 m}_{,k}\Lambda^{k\bar j},
 \label{ttt}
 \ee
 and with the same symmetry for the holomorphic indices. Here by $``,k"$ one
 means derivation with respect to the holomorphic coordinate $z^k$.
 The identity (\ref{ttt}) is satisfied {\em iff} the inverse of
 $\Lambda$ i.e. the two form $\omega$ is closed, namely,
 \be
 \omega_{i\bar j,k}=\omega_{k\bar j,i},\hspace{10mm} \omega_{i\bar j,\bar k}=\omega_{i\bar k,\bar j}.
 \ee
 Thus the K\"{a}hler transformation is given by closed 2-forms $K$ in (\ref{to}). An
 even dimensional Riemannian manifold can always be equipped with a complex structure for
 which the action functional,
 \be
 S\sim(d\omega,d\omega),
 \label{action}
 \ee
 is positive definite implying the desired condition,
 \be
 d\omega=0,
 \ee
 by the classical equation of motion.

 In summary the massless sector
 of the 2-from theory studied in the previous sections corresponds to the A
 string field theory.

  Here we slightly digress to
 discuss the difference between our method and that of \cite{sadov}
 where the theory of K\"{a}hler gravity is considered for the
 first time. The motivation in \cite{sadov} is as obtain a theory of
 K\"{a}hler transformation  to be the mirror of the Kodaira-Spencer
 theory of deformation of complex structure moduli. The so called AKS theory  of K\"{a}hler transformations is given by
 requiring that the operator
 \be
 D=d+[{d^c}^\dag,K],
 \label{tttt}
 \ee
 be nilpotent. On a manifold the operator $d$ can be defined with no
 reference to the complex structure. Given a K\"{a}hler structure by
 $\omega$ and the corresponding pull back operator $\Lambda$ one can
 define the operator ${d^c}^\dag$ by Eq. (\ref{tt}) which itself depends on $\omega$. Thus the AKS condition,
 \be
 D^2=0,
 \ee
 is well defined only by perturbation as far as ${d^c}^\dag$ in
 Eq. (\ref{tttt}) is  defined in terms of $d$ and
 $\omega=\omega_0+K$. In our approach one uses the  fact that the
 operator $d$ is defined independent of the K\"{a}hler structure
 and the equation of motion of the A string field theory is obtained from the
 condition,
 \be
 \left({d^c}^\dag\right)^2=[d,\Lambda]^2=0.
 \ee
 Thus our proposal is equivalent to the AKS conjecture if all terms in the
 perturbation are considered.

 There are some subtleties in defining the action (\ref{action}). The
 action $S$ in Eq. (\ref{action}),
 \be
 S\sim\int dK\wedge *d K,\label{hes}
 \ee
 is defined perturbatively since the Hodge $*$ operation is defined in terms of $\Lambda$
 which is known in terms of the deformation $K$ only via the following
 equation,
 \be
 \Lambda^{i\bar j}=\Lambda_0^{i\bar j}-\Lambda_0^{i\bar k} K_{k\bar k}\Lambda_0^{k\bar j}+\cdots.
 \ee
 Up to ${\cal O}(K^5)$ terms the action is given by as
 follows,
 \begin{eqnarray}
 S(K)\sim&&\int d V \left(K_{i\bar j,\bar k}K_{j\bar i ,k}+K_{j\bar i ,k}K_{i\bar j,\bar
 k}\right)\Lambda^{i\bar i}\left(\Lambda^{j\bar j}\Lambda^{k\bar k}-j\leftrightarrow
 k\right)\nn\\
 =&&\int d V \left(K_{i\bar j,\bar k}K_{j\bar i ,k}+K_{j\bar i ,k}K_{i\bar j,\bar
 k}\right)\times\nn\\
 &&\left\{\left(\Lambda_0^{i\bar i}\Lambda_0^{j\bar j}\Lambda_0^{k\bar k}\right)\right.\nn\\
 &&-\left(\Lambda_0^{i\bar l}K_{l\bar l}\Lambda_0^{l\bar i}\Lambda_0^{j\bar j}\Lambda_0^{k\bar k}+
 \Lambda_0^{i\bar i}\Lambda_0^{j\bar l}K_{l\bar l}\Lambda_0^{l\bar j}\Lambda_0^{k\bar k}
 +\Lambda_0^{i\bar i}\Lambda_0^{j\bar j}\Lambda_0^{k\bar l}K_{l\bar l}\Lambda_0^{l\bar
 k}\right)\nn\\
 &&+\left(\Lambda_0^{i\bar l}K_{l\bar l}\Lambda_0^{l\bar i}\Lambda_0^{j\bar m}K_{m\bar m}\Lambda_0^{m\bar j}\Lambda_0^{k\bar k}
 +\Lambda_0^{i\bar l}K_{l\bar l}\Lambda_0^{l\bar i}\Lambda_0^{j\bar j}\Lambda_0^{k\bar
 m}K_{m\bar m}\Lambda_0^{m\bar k}\right.\nn\\
 &&+\Lambda_0^{i\bar i}\Lambda_0^{j\bar l}K_{l\bar l}\Lambda_0^{l\bar j}\Lambda_0^{k\bar m}K_{m\bar m}\Lambda_0^{m\bar k}
 +\Lambda_0^{i\bar l}K_{l\bar l}\Lambda_0^{l\bar m}K_{m\bar m}\Lambda_0^{m\bar i}\Lambda_0^{j\bar j}\Lambda_0^{k\bar k}\nn\\
 &&+\left.\Lambda_0^{i\bar i}\Lambda_0^{i\bar l}K_{l\bar l}\Lambda_0^{l\bar m}K_{m\bar m}\Lambda_0^{m\bar j}\Lambda_0^{k\bar k}
 +\Lambda_0^{i\bar i}\Lambda_0^{j\bar j}\Lambda_0^{k\bar l}K_{l\bar l}\Lambda_0^{l\bar m}K_{m\bar m}\Lambda_0^{m\bar k}\right)\nn\\
 &&+\left.\mathcal{O}(K^5) -\left(j\leftrightarrow
 k\right)\right\}.
 \label{hesam}
 \end{eqnarray}
 The gauge condition,
 \be
 d^\dag K=0,
 \ee
 implies that, e.g.
 \be
 \Lambda_0^{i\bar j}K_{i\bar l,\bar j}+{\cal O}(K^2)=0.
 \ee
 Thus $K$ in this gauge is given by {\em transverse} modes, a
 well known notion in the ordinary quantum gravity.

 The action (\ref{hes}) is appeared to depend on the complex structure
 though the classical equation $d\omega=0$ was obtained with no
 reference to the complex structure.
 For simply connected manifolds, the independence of the action from the complex structure can be verified as
 follows.
 Any  variation of the complex structure deforms the operator $\bar{\rd}$ by
 $A\in H^{0,1}(TM)$ which is the Kodaira-Spencer field \cite{Vafa}. Since
 $A.K$ is a (0,2)-form, for a simply connected manifold
   it is vanishing identically,
 \begin{eqnarray}
 A.K&&=(A.K)_{\bar i\bar j}d\bar z^{\bar i}\wedge d\bar z^{\bar j}\nn\\
 &&=d\eta\nn\\
 &&=d(\eta_idz^i+\eta_{\bar i}dz^{\bar i})\nn\\
 &&=d(\eta_idz^i)+d(\eta_{\bar i}dz^{\bar i})\nn\\
 &&=d(d(\phi^{(1)}))+d(d(\phi^{(2)}))\nn\\
 &&=0,
 \end{eqnarray}
 where $\eta$ and $\phi^{(i)}$ are 1 and 0-forms on simply connected
 manifold respectively. In the 2nd and 5th lines we used the fact that for a simply connected $CY_3$,
 $h^{0,2}=h^{0,1}=0$.
 Consequently the theory is classically protected against complex structure
 deformation. In the next section we study the quantum background
 independence of this theory.

 Before closing this section we address the
 modification in the action (\ref{action}) for non-simply connected manifolds if it is used to describe the
 A string filed theory. The A model should be independent of the complex
 structure moduli. Thus one should add the condition
 \be
 A.K=0,
 \label{condition}
 \ee
 to the action using  the Lagrange multipliers $\alpha_i$,
 \be
 S=(d\omega,d\omega)+\left\{\alpha_i\int_{{\rm C}_i} A.K+h.c.\right\},
 \ee
 where C$_i$'s are $(0,2)$ cycles.
 The second term induces a coupling between the A-model and B$+\bar {\mbox{B}}$ model.\footnote{In \cite{td}, in a different approach to form field
 theories on $CY_3$,  it is argued that on a Calabi-Yau threefold, the A model is dual to the B$+\bar {\mbox{B}}$ model.}
  For a
 Calabi-Yau manifold the constraint (\ref{condition}) can be
 equivalently given by
 \be
 A'\wedge K=0.
 \ee
  Thus the modified action can be given by
 \be
 S=(d\omega,d\omega)+\left\{\int_{{\rm CY_3}}\lambda\wedge A'\wedge
 K+h.c.\right\},\hspace{1cm}\lambda\in H^{0,1}.
 \ee
 We postpone further investigation of this theory to future works.

 \section{Quantum background independence}
 The observables of the A string field theory should be independent of
 the  complex structure.
 In this section at  first we construct the necessary tools to study the theory of complex structure
 deformation,
 then  we use these tools to study the quantum background independence of
 $H^2(CY_3)$ and $H^3(CY_3)$.
 The quantum background independence of
 $H^3(CY_3)$ is worked out previously and is  known to  result in the holomorphic anomaly
 equations \cite{Witten1,Verlinde,Gerasimov,Loran}.
\subsection{The structure of connections}

In this subsection we study the connections corresponding to the
deformation of the complex structure on a complex manifolds. It
is known  that the number of moduli complex structure is
$h^{2,1}$. An infinitesimal deformation of complex structure makes
the holomorphic differentials $dz^i$ on the complex manifold to be
mixed  with the anti-holomorphic differentials \cite{Strominger},
\begin{equation}
\label{id} \frac{\rd}{\rd t_M} dz^i=\nu_{j M}^i
dz^j+\mu_{\overline{j} M}^i d\overline{z}^{\overline{j}},
\end{equation}
where $t_M$ are variables of complex structure moduli space and
$M$ counts the number of complex structure moduli,
$M=1,\cdots,h^{21}$. Fundamental duality relation between basis
of forms and basis of vectors,
\begin{eqnarray}
\label{bid} &&<dz^i,\overline{\partial}_{\overline{j}}>=
<d\overline{z}^{\overline{i}},\partial_j>=0,\nn\\
&&<dz^i,\partial_j>=\delta^i_j,\nn\\
&&<d\overline{z}^{\overline{i}},\overline{\partial}_{\overline{j}}>=
\delta^{\overline{i}}_{\overline{j}},
\end{eqnarray}
 should be satisfied  after
the deformation. Therefore we obtain the following relations for
those basis,
\begin{eqnarray}
\label{bit} &&\frac{\rd}{\rd t_M} dz^i=\nu_{j M}^i
dz^j+\mu_{\overline{j}M}^i
d\overline{z}^{\overline{j}},\nn\\
&&\frac{\rd}{\rd t_M} \partial_{i}=-\nu_{i M}^j
\partial_j+\epsilon^{\overline{j}}_{i M}\overline{\partial}_{\overline{j}},\nn\\
&&\frac{\rd}{\rd t_M}
d\overline{z}^{\overline{i}}=\xi_{\overline{j}
M}^{\overline{i}}d\overline{z}^{\overline{j}}-\epsilon_{j M}^i dz^j,\nn\\
&&\frac{\rd}{\rd t_M}
\overline{\partial}_{\overline{{i}}}=-\xi_{\overline{i}
M}^{\overline{j}}
\overline{\partial}_{\overline{j}}-\mu^{\overline{j}}_{\overline{i}
M}\overline{\partial}_{\bar j}.
\end{eqnarray}
Thus we have four sets of independent connections $\mu_M$,
$\nu_M$, $\epsilon_M$, and $\xi_M$ of complex structure
deformation on a complex manifold. In the next subsection we show
on a $CY_3$ manifold only two sets of these connections remain
and two other sets are vanishing identically.
\subsection{Complex structure independence in form field theories}

In the 3-from gravity theory, which is defined on a Calabi-Yau
threefold manifold, real closed 3-form $\gamma_0$ can be
decomposed in the basis consisting of the Calabi-Yau 3-form
$\Omega\in H^{3,0}$, \footnote{For Calabi-Yau threefold we have
$h^{0,0}=h^{3,0}=h^{0,3}=h^{3,3}=1$} its covariant derivatives
 with respect to the moduli directions $t_M$,  $\nabla_M\Omega\in H^{2,1}$,\footnote{$H^3(CY_3)$ forms a bundle over the moduli
space of the complex structures of the $CY_3$. The holomorphic
3-form defines a line sub-bundle $\mathcal{L}$ and on this
sub-bundle one can define a connection ${\nabla_M=\rd_M+\rd_MK}$
where $K$ is the K\"{a}hler potential  defined as follows,
\begin{equation}
K=-\ln i\int\bar{\Omega}\wedge\Omega.\nn
\end{equation}
  }
 and their complex conjugates,
\begin{equation}
\gamma_0=\lambda^{-1}\Omega+x^M\n_M\Omega+\bar x^{\bar
M}\bar\n_{\bar M}\bar\Omega+\bar\lambda^{-1}\bar\Omega.
\label{gamma}
\end{equation}
Locally the Calabi-Yau 3-form takes the form,
\begin{equation}
\Omega=\frac{1}{3!}h(z)\varepsilon_{i j k}dz^i\wedge dz^j\wedge
dz^k,
\end{equation}
and
\begin{equation}
\frac{\rd}{\rd t_M}\Omega=\frac{1}{3!}h(z)_{,M}\varepsilon_{i j
k}dz^i\wedge dz^j\wedge dz^k+\frac{1}{2!}h(z)\varepsilon_{i j
k}\frac{\rd}{\rd t_M}(dz^i)\wedge dz^j\wedge dz^k.
\end{equation}
By expanding this in terms of $\mu_{\bar{j}M}^i$ and $\nu_{j M}^i$
and using Eqs. (\ref{bit}) one verifies that $\mu_{\bar{j}M}^i$ and
$\nu_{j M}^i$ are generators  of bidegree (2,1) and (3,0)-forms
respectively.

Since on a Calabi-Yau threefold $\rd_M\bar{\Omega}=0$ and $\Omega$
is a nowhere vanishing holomorphic 3-form, by expanding
$\rd_M\bar{\Omega}$ one can verify that
$\xi_{\bar{b}M}^{\bar{a}}=0=\epsilon_{b M}^a$ and Eqs.(\ref{bit})
simplify to
\begin{eqnarray}
 &&\frac{\rd}{\rd t_M} dz^i=\nu_{j M}^i dz^j+\mu_{\bar{j}M}^i
d\bar{z}^{\bar{j}},\nn\\
&&\frac{\rd}{\rd t_M} \partial_{i}=-\nu_{i M}^j
\partial_j,\nn\\
&&\frac{\rd}{\rd t_M} d\bar{z}^{\bar{i}}=0,\nn\\
&&\frac{\rd}{\rd t_M}
\bar{\partial}_{\bar{{i}}}=-\mu^{\bar{j}}_{\bar{i}
M}\bar{\partial}_{\bar j}.
\end{eqnarray}
Consequently  all of the complex structure deformation on
Calabi-Yau threefold are given by  $\mu_{\bar{b}M}^a$ and $\nu_{b
M}^a$.

In 3-form theory it is shown that the holomorphic anomaly
equations depend only on the  $\mu_{\bar{b}M}^a$ connections
\cite{Vafa1} which can be equivalently obtained by imposing the
background independence condition on $H^3(CY_3)$
\cite{Verlinde,Loran}.

For the  2-form gravity theory one deals with  a 2-form
$\omega_0$ on the $CY_3$. The  2-form $\omega_0$ is given in
terms of a  basis of $H^2(CY_3)$ as follows,
\begin{equation}
\omega_0=x_I\omega^I(z),
\end{equation}
where $I$ runs over the  basis of $H^2(CY_3)$. If a compact simply
connected $CY_3$ manifold is considered, since for any suitable
complex structure we have $h^{2,0}=0$ any basis for $H^2(CY_3)$ is
a basis for the K\"{a}hler forms. For a given complex structure
$\omega^I(z)=\omega^I(z)_{a \bar{b}}dz^a\wedge d\bar{z}^{\bar{b}}$
for $I=1,\cdots h^{1,1}$. Thus one obtains,
\begin{equation}
\rd_M\omega^I=(\omega^I_{i\bar{j}})_{,M}dz^i\wedge
d\bar{z}^{\bar{j}}+\omega^I_{i \bar{j}}\rd_M(dz^i)\wedge
d\bar{z}^{\bar{j}}.
\end{equation}
By expanding this formula in terms of $\mu_{\bar{j}M}^i$ and
$\nu_{j M}^i$ and noting that $h^{2,0}(CY_3)=0$ one obtains that
$\mu_{\bar{j}M}^i$ is a null operator on the $H^2$ i.e.
 \be
 \mu_{\bar{i} M}^i\omega_{i
 \bar{j}}d\bar{z}^{\bar{i}}\wedge d\bar{z}^{\bar{j}}=0.
 \ee
 By imposing the background (i.e. complex structure) independence
condition on the symplectic 2-form $\omega_0$, one obtains,
\begin{equation}
[\nu_{M i}^j\omega^I_{j
\bar{i}}+(\omega^I_{i\bar{i}})_{,M}]dz^i\wedge d\bar{z}^{\bar{i}}=0.
\end{equation}
Thus, $\nu_{M i}^j$ are generators of transformation along the
complex structure moduli directions for $H^2(CY_3)$.


\appendix
\section{Abelian duality in
7 dimensions }

At the end of section 2 we saw that a coefficient multiplies the
partition functions  of 2-form  theory in 7 dimensions. In this
appendix we  show that this coefficient is a topological
invariant which only depends on the Hodge numbers of $CY_3$. For
deriving that coefficient one should calculate two terms; the
Gaussian integral over $A'$, $\int \mathcal{D}A'
e^{-\frac{\Lambda}{4\pi}\int_\Sigma A'\wedge\ast_7A'}$, and the
grassmanian integral over $c$ and $\bar{c}$, $\int
\mathcal{D}\bar{c} \mathcal{D}c
e^{-\frac{\Lambda}{4\pi}\int_\Sigma\bar{c}c}$. For the first one,
we note that $A'$ is a  4-form and  the Hodge star operator is a
linear map. Thus this term gives a factor of $\int dy
e^{-(\Lambda/4\pi)y}=\sqrt{\frac{\pi}{\Lambda/4\pi}}$ for each
4-form.\footnote{ $y$ is a normalization real variable and is
defined such that  $\int\mathcal{D}A'e^{-\int_7A'\wedge\ast
A'}=\int dy e^{-y}$ \cite{qf}.}

For the second term it should be noted that the path integral
over the ghost fields run over the set of grassmanian $r$-forms
and $(7-r)$-forms where $r=0,\dots ,7$ (note that either  the $r$
or ($7-r$)-form is an even form). For each set of ghost fields one
obtains $\int\mathcal{D}\bar{c}\mathcal{D}c
e^{-\frac{\Lambda}{4\pi}\int \bar{c}c }=\Lambda/4\pi$. Thus the
overall factor is
\begin{equation}
\mathcal{C}(\Sigma)=(2\pi)^{1+h^{2,1}+h^{1,1}/2}(\Lambda/2\pi)^{4+h^{2,1}+3h^{1,1}/2},
\end{equation}
where we have used the following identities for the Hodge numbers
of $\Sigma=CY_3\times S^1$ \cite{Nakahara},
\begin{eqnarray}
&&b^0(\Sigma)=b^7(\Sigma)=1,\nn\\
&&b^1(\Sigma)=b^6(\Sigma)=2,\nn\\
&&b^2(\Sigma)=b^5(\Sigma)=h^{1,1}(CY_3),\nn\\
&&b^3(\Sigma)=b^4(\Sigma)=2h^{2,1}(CY_3)+h^{1,1}(CY_3)+2.
\end{eqnarray}

 \section{Relation between the operators $d$ and $\ast$ in $m$ and $m+1$ dimensions} In
 this appendix we deal with some relations between the operators in
 $m$ and $m+1$ dimensions. In section 2 we used two elementary
 operators, namely the  exterior derivative  $d$ and the Hodge star
 $\ast$ in 6 and 7 dimensions.

 Operator $d$ is a map $d:\Omega^r(M)\rightarrow\Omega^{r+1}(M)$, where
 $\Omega^r(M)$ are r-forms of  manifold $M$, whose action on a
 r-form,
 \begin{equation}
 \eta_r=\frac{1}{r!}\eta_{\mu_1\cdots\mu_r}dx^{\mu_1}\wedge\cdots\wedge
 dx^{\mu_r},
 \end{equation}
 is defined by,
 \begin{equation}
 d\eta_r=\frac{1}{r!}(\frac{\partial}{\partial
 x^\nu}\eta_{\mu_1\cdots\mu_r})dx^\nu\wedge
 dx^{\mu_1}\wedge\cdots\wedge dx^{\mu_r}.
 \end{equation}
 To obtain the relation between exterior derivative in $m$ and
 $m+1$ dimensional manifolds, one should note that for our purpose
 $(m+1)$-th direction in $(m+1)$-dimensional manifold is
 perpendicular to other directions. Therefore it is obvious that,
 \begin{equation}
 d_{m+1}=d_m + dt\wedge\frac{\partial}{\partial t},
 \end{equation}
 where subscripts $m$ and $m+1$ indicate the dimension of manifold
 and $t$ is the coordinate of $(m+1)$-th direction.

For Hodge star operator there are some subtleties. Operator $\ast$
is a linear map $\Omega^{r}(M)\rightarrow\Omega^{m-r}(M)$, where
$m$ is the dimension of manifold $M$. The operation of the Hodge
$*$ on an element  of $\Omega^{r}(M)$ is defined by,
\begin{equation}
\ast(dx^{\mu_1}\wedge\cdots\wedge
dx^{\mu_r})=\frac{\sqrt{|g|}}{r!(m-r)!}g^{\mu_1 \nu_1}\cdots
g^{\mu_r \nu_r}
\varepsilon_{\nu_1\cdots\nu_r\nu_{r+1}\cdots\nu_m}dx^{\nu_{r+1}}\wedge\cdots\wedge
dx^{\nu_m},
\end{equation}
in which  $g^{\mu \nu}$ is the metric of manifold $M$, $g$ is the
determinant of $g_{\mu \nu}$, and $\mu_i , \nu_i=1,\cdots ,m$.

We consider two independent cases; an $r$-form in $m+1$ dimensions
which is an $r$-form along $m$-dimensional manifold, and an
$r$-form in $m+1$ dimensions which is an $(r-1)$-form along
$m$-dimensional manifold and an 1-form along the $(m+1)$-th
direction. For the first case
$\eta_r=\frac{1}{r!}\eta_{\mu_1\cdots\mu_r}dx^{\mu_1}\wedge\cdots\wedge
dx^{\mu_r}$, thus,
\begin{eqnarray}
&&\ast_m\eta_r=\frac{\sqrt{|g|}}{r!(m-r)!}\eta_{\mu_1\cdots\mu_r}g^{\mu_1
\nu_1}\cdots g^{\mu_r \nu_r}
\varepsilon_{\nu_1\cdots\nu_r\nu_{r+1}\cdots\nu_m}dx^{\nu_{r+1}}\wedge\cdots\wedge dx^{\nu_m},\label{ast m}\\
&&\ast_{m+1}\eta_r=\frac{\sqrt{|g'|}}{r!(m+1-r)!}\eta_{\mu_1\cdots\mu_r}g^{\mu_1
\nu_1}\cdots g^{\mu_r \nu_r}
\varepsilon_{\nu_1\cdots\nu_r\nu_{r+1}\cdots\nu_m \nu_{m+1}}\nn\\
&&\hspace{74mm}dx^{\nu_{r+1}}\wedge\cdots\wedge dx^{\nu_m}\wedge
dx^{\nu_{m+1}},\label{farhang}
\end{eqnarray}
where $g$ and $g'$ are determinant of metrics in $m$ and $m+1$
dimensional manifolds respectively, $\varepsilon_{1\cdots m}$ is
the totally antisymmetric tensor, ${\mu_i=1,\cdots, m}$,
${\nu_i=1,\cdots, m}$  for the first equation, and
${\nu_i=1,\cdots, m}$ for the second equation. Since the
$(m+1)$-th direction is perpendicular to the other directions one
obtains $g^{\mu_i,m+1}=\delta_{\mu_i,m+1}$ and $g=g'$. Therefore
in Eq. (\ref{farhang}) $\nu_i$ only takes value in the range
$1\le\nu_i\le m$ for $i=,1\cdots, r+1$. Since there are $(m+1-r)$
positions for $\nu_{m+1}$ in of $\varepsilon_{1\cdots m}$, one
obtains,
\begin{eqnarray}
\ast_{m+1}\eta_r=\frac{\sqrt{|g|}}{r!(m-r)!}\eta_{\mu_1\cdots\mu_r}g^{\mu_1
\nu_1}\cdots g^{\mu_r \nu_r}
\varepsilon_{\nu_1\cdots\nu_r\nu_{r+1}\cdots\nu_m
{m+1}}\nn\\
 dx^{\nu_{r+1}}\wedge\cdots\wedge dx^{\nu_m}\wedge
dx^{m+1},\label{ast m1}
\end{eqnarray}
where $\mu_i , \nu_i=1\cdots m$. By comparing Eq. (\ref{ast m})
with Eq. (\ref{ast m1}) one obtains the following  relation,
\begin{equation}
\ast_{m+1}\eta_r=\ast_m\eta_r\wedge dx^{m+1}.\label{m}
\end{equation}

In the second case, we assume,
\begin{equation}
\eta_{r+1}=\frac{1}{(r+1)!}\eta_{\mu_1\cdots\mu_r
m+1}dx^{\mu_1}\wedge\cdots\wedge dx^{\mu_r}\wedge dx^{m+1}
=\frac{1}{r+1}\eta'_{r m+1}\wedge dx^{m+1},
\end{equation}
where $\eta'_{r m+1}\equiv\frac{1}{r!}\eta_{\mu_1\cdots\mu_r
m+1}dx^{\mu_1}\wedge\cdots\wedge dx^{\mu_r}$ is an $(r+1)$-form
along the $(m+1)$-dimensional manifold and an $r$-form along the
$m$-dimensional manifold. Hodge star operation in the $m$-manifold
gives,
\begin{eqnarray}
\ast_{m}\eta_{r+1}&&=\frac{1}{r+1}\ast_m(\eta'_{r m+1})\wedge dx^{m+1}\nn\\
&&=\frac{1}{r+1}\frac{\sqrt{|g|}}{r!(m-r)!}\eta_{\mu_1\cdots\mu_r
m+1}g^{\mu_1 \nu_1}\cdots g^{\mu_r \nu_r}
\varepsilon_{\nu_1\cdots\nu_r\nu_{r+1}\cdots\nu_m}\nn\\
&&\hspace{50mm}dx^{\nu_{r+1}}\wedge\cdots\wedge dx^{\nu_m}\wedge
dx^{m+1},\label{b1}
\end{eqnarray}
where $\mu_i , \nu_i=1\cdots m$. In addition,
\begin{eqnarray}
\ast_{m+1}\eta_{r+1}&=&\frac{\sqrt{|g'|}}{(r+1)!(m-r)!}
\eta_{\mu_1\cdots\mu_r m+1}\nn\\
&&\hspace{19mm}g^{\mu_1 \nu_1}\cdots g^{\mu_r \nu_r}g^{m+1 \nu}
\varepsilon_{\nu_1\cdots\nu_r\nu\nu_{r+2}\cdots\nu_{m+1}}
dx^{\nu_{r+2}}\wedge\cdots\wedge dx^{\nu_{m+1}}\nn\\
&=&\frac{\sqrt{|g|}}{(r+1)!(m-r)!}
\eta_{\mu_1\cdots\mu_r m+1}\nn\\
&&\hspace{7mm}g^{\mu_1 \nu_1}\cdots g^{\mu_r \nu_r}g^{m+1, m+1}
\varepsilon_{\nu_1\cdots\nu_r,m+1,\nu_{r+2}\cdots\nu_{m+1}}
dx^{\nu_{r+2}}\wedge\cdots\wedge dx^{\nu_{m+1}}\nn\\
&=&(-)^{m-r}\frac{\sqrt{|g|}}{(r+1)!(m-r)!}
\eta_{\mu_1\cdots\mu_r m+1}\nn\\
&&\hspace{17mm}g^{\mu_1 \nu_1}\cdots g^{\mu_r \nu_r}
\varepsilon_{\nu_1\cdots\nu_r\nu_{r+2}\cdots\nu_{m+1} m+1}
dx^{\nu_{r+2}}\wedge\cdots\wedge dx^{\nu_{m+1}}.\label{b2}
\end{eqnarray}
Comparing Eq. (\ref{b1}) with Eq. (\ref{b2}) one verifies that
\begin{equation}
(\ast_{m+1}\eta_{r+1})\wedge
dx^{m+1}=(-)^{(m-r)}\ast_m\eta_{r+1}.\label{m1}
\end{equation}
If the signature of  $(m+1)$-th direction is  Lorentzian, then
$g^{\mu_i,m+1}=-\delta_{\mu_i,m+1}$ and one should consider an
extra $(-1)$ in the above equation.

 As an application of these relations one
 can show that for an $r$-form $\eta_r$ along 6 dimensions of
 original 7-dimensional manifold,
 \bea
 d^\dag_7d_7\eta_r&&=(-)^{7(r+2)+1}*_7d_7*_7d_7\eta_r\nn\\
 &&=(-)^{r+1}*_7d_7*_7\left(d_6\eta_r+dt\wedge\dot{\eta_r}\right)\nn\\
 &&=(-)^{r+1}*_7d_7\left(*_6d_6\eta_r\wedge
 dt+*_6\dot{\eta_r}\right)\nn\\
 &&=(-)^{r+1}*_7\left(d_6*_6d_6\eta_r\wedge
 dt+d_6*_6\dot{\eta_r}+dt\wedge*_6\ddot{\eta_r}\right)\nn\\
 && =(-)^{r+1}\left((-)^r*_6d_6*_6d_6\eta_r+*_6d_6*_6\dot{\eta_r}\wedge dt+*_6*_6\ddot{\eta_r}\right)\nn\\
 &&=d^\dag_6d_6\eta_r-\ddot{\eta_r}+(-)^rd^\dag_6\dot{\eta_r}\wedge
 dt,\label{ddagad}
 \eea
 where Eq. (\ref{dag1},\ref{ast}) are used to this derivation.
 Also  for $\eta_{r+1}=\eta_r\wedge
 dt$, one obtains,
 \bea
 d^\dag_7d_7(\eta_r\wedge dt)&&=(-)^{7(r+2)+1}*_7d_7*_7d_7(\eta_r\wedge dt)\nn\\
 &&=(-)^{r+1}*_7d_7*_7\left(d_6\eta_r\wedge dt\right)\nn\\
 &&=*_7d_7\left(*_6d_6\eta_r\right)\nn\\
 &&=*_7\left(d_6*_6d_6\eta_r+dt\wedge *_6d_6\dot{\eta_r}\right)\nn\\
 && =*_6d_6*_6d_6\eta_r\wedge dt+*_6*_6d_6\dot{\eta_r}\nn\\
 &&=-d^\dag_6d_6\eta_r\wedge dt+(-)^{r+1}d_6\dot{\eta_r}.\label{ddaga}
 \eea

\end{document}